# Distinctive Thermoelectric Properties of Supersaturated Si-Ge-P Compounds: Achieving Figure of Merit ZT > 3.6

Swapnil Ghodke,[†]* Omprakash Muthusamy,[†]* Kévin Delime–Codrin,[†] Seongho Choi,[†] Saurabh Singh,[†] Dogyun Byeon,[†] Masahiro Adachi,[‡] Makoto Kiyama,[‡] Takashi Matsuura,[‡] Yoshiyuki Yamamoto,[‡] Masaharu Matsunami,[†] Tsunehiro Takeuchi[†]

[†]Energy Materials Laboratory, Toyota Technological Institute, 2-12-1 Hisakata, Tempaku Ward, Nagoya, 468-0034, Japan

[‡]Sumitomo Electric Industries Ltd., Itami, Hyogo 664-0016, Japan.

ABSTRACT

The efficiency of energy conversion in thermoelectric generators (TEGs) is directly proportional to electrical conductivity and Seebeck coefficient while inversely to thermal conductivity. The challenge is to optimize these interdependent parameters simultaneously. In this work, the problem is addressed with a novel approach of nanostructuring and constructive electronic structure modification to achieve a very high-value of dimensionless figure-of-merit $ZT > 3.6$ at 1000 K with negative Seebeck coefficient. Supersaturated solid-solutions of Si-Ge containing ~1 at.% Fe and 10 at.% P are prepared by high-energy ball milling. The bulk samples consisting of ultra-fine nano-crystallites (9.7 nm) are obtained by the sophisticated "low-temperature & high-pressure sintering process." Despite that the electrical resistivity is slightly high due to the localization of electrons is associated with the highly disordered structure and low electrical density of states near the chemical potential, a very low thermal conductivity $\kappa$ less than 1 W m$^{-1}$K$^{-1}$ and very large magnitude of Seebeck coefficient $|S|$ exceeding 470 µV K$^{-1}$ are achieved in association with the nanostructuring and the Fe $3d$ impurity states, respectively, to realize a very large magnitude of $ZT$.

Introduction

The electrical energy plays as an essential part of modern human society. The energy is generated by consuming nonrenewable resources, which on the contradictory is also responsible for the releasing greenhouse gasses in the atmosphere. Greenhouse gases can hamper the ecological balance by disrupting the ecosystem through global warming and climate change. The imbalance in demand and supply of these limited energy resources also forecasts a global energy crisis for the future generation. The solution to the above problems lies in alternative energy resources and/or new technologies with higher efficiencies of energy conversion. Here, the TEG, in which waste heat is converted into useful electrical energy, is emphasized as one of the potential technologies to reduce the carbon footprint and utilize the energy resources more efficiently. The efficiency energy conversion in TEGs is an increasing function of the dimensionless figure-of-merit, $ZT = S^2 \sigma T \kappa^{-1}$, where $S$, $\sigma$, $T$, and $\kappa$ stand for the Seebeck coefficient, electrical conductivity, absolute temperature, and thermal conductivity of constituent thermoelectric materials, respectively.[1,2]

Since early 1900s, extensive research has been carried out for obtaining high-performance thermoelectric materials for practical applications.[3] Numerous state-of-art thermoelectric materials have been reported with $ZT$ more than unity; SnSe ($ZT$ = 2.6),[4,5] Pb–Te ($ZT$ ~ 1.8),[6] Bi–Sb–Te ($ZT$ ~ 1.86),[7,8] Zn–Sb ($ZT$ ~ 1.3),[9] $Cu_2Se$ ($ZT$ > 2),[10,11] $Mg_2Si$ ($ZT$ ~ 1.3),[12] TAGS ($ZT$ ~ 1.5),[13] $MnSi_\gamma$ ($ZT$ ~ 1.15)[14–16]. Unfortunately, the drawbacks of these compounds are either (a) Usage of toxic or expensive constituent elements or (b) Thermal degradation of transport properties, and/or (c) Low $ZT$ for commercial application, where standard requirement is $ZT$ > 2.5.[4,17–20] Theoretically, the probability of achieving such high $ZT$ > 2.5 would need a large temperature gradient, and that is attainable only with the high-temperature thermoelectric materials.

Silicon-germanium (Si–Ge) alloys are widely used in photovoltaic,[21] transistors,[22] optoelectronic-photonic devices,[23] and have investigated into the thermoelectric properties ($ZT$ ~ 1.9)[24–27] for high-temperature applications[28–32]. Si–Ge based alloys are definitely nontoxic, environmentally friendly. In the application point of view, Si–Ge alloys are characterized by high mechanical strength, low coefficient of thermal expansion mismatch, high-melting-point, and scalable with the silicon technology.[33–35]

NASA reported a decent figure-of-merit 0.5 (*p*-type) and 0.9 (*n*-type) in the 1970s [36] to utilize them in radioisotope thermoelectric generators. Since then, a large number of investigations reported the improvement in thermoelectric properties by reducing lattice thermal conductivity through grain boundary scattering,[37–41] nanostructuring,[25–27,42–45] composite effect,[46–50] quantum dot superlattices,[51,52] amorphous structures,[53,54] modulation doping,[55,56] and nanoporous structures.[57]

Recently, we succeeded in developing an *n*-type Si-Ge-P-Fe sample possessing $ZT = 1.88$ at 900 K.[24] High figure-of-merit was achieved by constructive modification in the electronic structure through impurity states of Fe 3*d* and nanostructuring by high-energy ball milling. However, the investigation into their thermal stability was limited up to 900 K. Besides, the negative temperature coefficient of electrical resistivity (TCR) was quite unusual for a heavily doped degenerated semiconductor. Furthermore, there was also a possibility to observe hysteresis in temperature-dependent Seebeck coefficient, electrical resistivity, and thermal conductivity due to phase transition, while thermal (heating-cooling) cycling. All the concerns mentioned above and the unexplained problems led us to carry out the current investigation.

This paper describes the formation of supersaturated Si-Ge-P-Fe alloys with high energy ball milling, the thermal stability of such metastable phases, and phase transition/ crystallization/ impurity precipitation at high-temperatures. Then, the microstructure analysis of the amorphous-like nanostructure, which was responsible for the ultra-low thermal conductivity. Furthermore, we report the temperature-dependent thermoelectric properties with adequate reasoning for unusual behavior observed for Seebeck coefficient, electrical resistivity, and thermal conductivity in the heating-cooling cycles. Finally, we report an extremely high value of $ZT > 3.6$ obtained for the Si-Ge-P-Fe bulk samples.

EXPERIMENTAL PROCEDURES

High purity Si (99.99%), Ge (99.99%), and P (99.9999%) chunks in a stoichiometric ratio of $Si_{55}Ge_{35}P_{10}$ were ground and mixed thoroughly with a mortar and pestle in a glove box under argon atmosphere. Further, the powder was sealed in stainless steel mill pot with 20: 1 ball-to-powder ratio. The ball milling was carried out in two steps: (Step I) alloying phase, where powder was milled at 600 rpm for 6 – 10 h to form supersaturated solid-solution of Si-Ge-P, (Step II) grinding phase, to obtain ultra-fine nano-grains the powder was milled at 300 rpm for 50 h with a cycle of 30 min milling and 15 min pause time. Pure Si and Ge are quite susceptive for the oxidation[86], to prevent the oxidation during the milling; the mill pot was refilled at the periodic interval with a mixed gas of argon and hydrogen.

The alloyed powder was collected from the mill pot and filled into a tungsten carbide die ($\phi$ = 10 mm) in the argon atmosphere of the glove box. The powder was then sintered by Spark Plasma Sintering (SPS) technique with sophisticated sintering parameters, i.e., low-temperature (T = 823 – 873 K) & high-pressure (P = 400 – 600 MPa) with a very long soaking time (4 h). In this study, we have analyzed three samples for investigating the effect of milling parameters, (a) SM#1/ reference data: sintered after alloying and grinding phase (Step I + II) measurements up to 873 K,[24] (b) SM#2: sintered after alloying and grinding phase (Step I + II) measurements up to 1173 K, and (c) SM#3: sintered after alloying phase (Step I) measurements up to 1173 K. The density of the bulk sample was measured by the Archimedes method in ethanol media.

The thermal stability of the metastable supersaturated solid-solution was investigated by differential thermal analysis (DTA) using RIGAKU TG8121 in the temperature range of 300 K – 1173 K with the heating rate of 10 K/min under argon atmosphere. The phases involved in each sample were analyzed by a conventional powder X-ray diffraction (XRD) using Cu–K$\alpha$ radiation ($\lambda$= 0.15418 nm) in Bruker D8 Advance. The chemical composition was investigated by energy-dispersive X-ray spectrometry (EDX) from JEOL JED-2140GS equipped with a scanning electron microscope (SEM) JEOL JSM-6330F operated at an accelerating voltage of 15 kV. For the transmission electron microscopic (TEM) analysis, the synthesized Si-Ge-Fe-P powder was grounded in ethanol solution to remove agglomeration of particles. Some drops of the resulting dispersion were dropped on the copper grids with holey carbon films. The grid was examined in a JEM-2100 (JEOL, Japan) operated at 200 kV under vacuum ($2.0 \times 10^{-5}$ Pa).

The thermal diffusivity ($D$) of the sintered pellet was measured in the temperature range of 300 – 1173 K with using Laser flash method (NETZSCH LFA457), specific heat ($C_p$) was extracted from the reference sample (Pyroceram9606). Electrical resistivity and Seebeck coefficient were measured using standard four-probe technique and steady-state method, respectively, in the same temperature range (300 – 1173 K) and heating – cooling rate 160 K/h under a vacuum condition ($10^{-2}$ Pa). The error bar or the uncertainty for the Seebeck coefficient (7 %), electrical resistivity ±5 %, thermal conductivity (5 – 7 %) and the uncertainty for $ZT$ was about ±10 %. The uncertainty in the transport properties was obtained from several measurements and instrumental inaccuracy.

RESULTS

SUPERSATURATED SI-GE-P PHASE STABILIZATION

The supersaturated metastable phase of Si-Ge-P-Fe in ball-milled powder (Step I) was investigated for its thermal stability by heating the sample to 1173 K in DTA measurement. A small exothermic peak was observed at 916 K in the DTA curve shown in **Figure 1a**. This fact indicates that a phase transition or crystallization of the metastable phase takes place in limited portions of the sample.

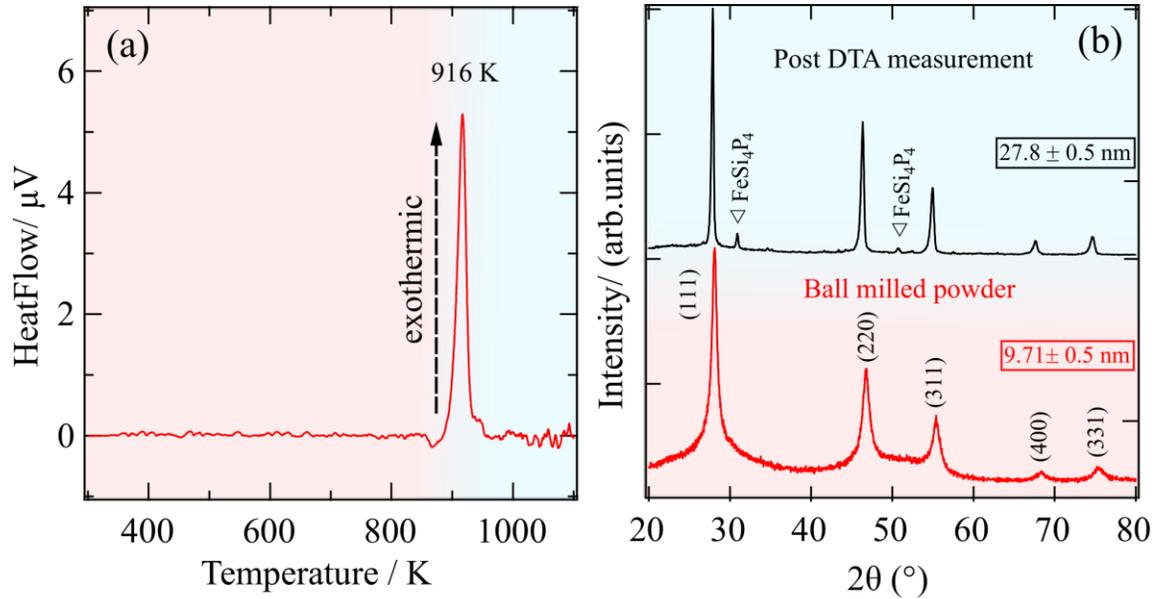

**Figure 1.** (a) The DTA measurement of step (I + II) ball milled powder: supersaturated Si-Ge-Fe-P phase and (b) XRD patterns measured for before and after DTA measurement.

The XRD patterns of ball-milled powder before and after the DTA measurement are plotted in **Figure 1b**. The dissimilarity in the two patterns is noticeable. The XRD pattern of the ball-milled powder represents a typical Si-Ge cubic structure (space group: *Fd-3m* (227)) with broadening in the peaks due to the formation of nanostructures. Presence of amorphous phase was confirmed in halo pattern in background intensity, and also an absence of any impurity peaks. Formation of a supersaturated solution of Si-Ge through the mechanical alloying process was consistent with the reports of Zhang *et al.* and Suryanarayana *et al.*[33,58,59] Whereas, the XRD pattern of heat-treated powder showed significant sharpening of the peaks, most probably due to improved crystallinity and grain growth. The halo pattern was disappeared after the heating, and the small peaks corresponding to $FeSi_4P_4$ precipitation were appeared. Similar impurity peaks were observed in our previous study[24] on the sample sintered at 1173 K above the phase transition temperature. Hence the precipitation of $FeSi_4P_4$ from the supersaturated phase is conclusive and very consistent.

To understand the transformation in crystallinity with heat treatment the crystallite sizes was calculated from full-width-half-maximum (FWHM) by using Scherrer's formula $D = K\lambda/\beta cos(\theta)$, Here $K$, $\lambda$, $\beta$, and $\theta$ represent Debye constant ($K = 0.94$), X-ray wavelength ($\lambda = 0.15418$ nm), line broadening, and Bragg angle, respectively. The crystallite size for the as the milled powder was $9.7 \pm 0.5$ nm, and that for the post-DTA powder $27.8 \pm 0.5$ nm. The drastic increase in crystallite size could be related to the crystallization of the amorphous phase accompanied by the precipitation of secondary phase. If it is the case, we need to synthesize bulk sample below the transition temperature for maintaining a stable supersaturated solid-solution with an ultrafine nanostructure. Therefore, tailored sintering condition of high-pressure (400 MPa), low-temperature (873 K) with long soaking time (4 h) was employed to obtain a dense bulk pellet.

The x-ray diffraction patterns of supersaturated Si-Ge-Fe-P ball milled powder and the sintered pellet were accumulated at room temperature and plotted together with the calculated patterns of constituent elements in **Figure S1**. We succeeded in obtaining a dense bulk sample without any precipitation of secondary phases. Although the halo pattern in the XRD was disappeared, the peak broadening was retained due to tailored sintering condition, and also the crystallite size was slightly increased to $15.8 \pm 0.5$ nm.

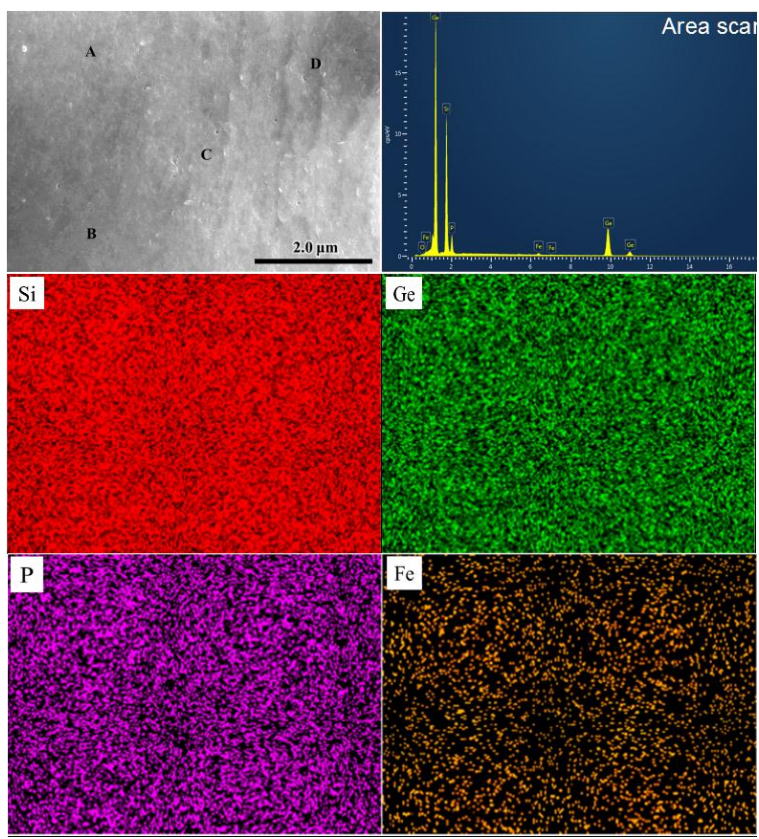

**Figure 2.** SEM image (top) shows the homogeneous microstructure of the SM#3 sample (fractured surface). A, B, C, and D represent positions for the EDX analysis. Elemental mapping (bottom) for Si, Ge, P, and Fe elements.

Microstructure and chemical composition of the ball-milled powder (**Figure S2**) and the sintered pellet (**Figure 2**) were analyzed by SEM-EDX. The milled powder was consisting of particles in sub-micron sizes, while at higher magnification; a very fine microstructure of nano-grains was observed, as shown in **Figure S2a – S2b**. We also carry out elemental mapping on the powder to qualitatively analyze the homogeneity in the composition. The mapping shows a desired composition or distribution of Si, Ge, P, and Fe (from the stainless steel mill pot and balls). The detailed SEM-EDX for qualitative & quantitative analysis were done precisely by elemental mapping, point EDX, and an area scan on fractured surface of the sintered sample which is shown in **Figure 2**. The obtained chemical compositions are summarized in **Table 1**.

**Table 1**. Chemical composition of **SM#3** bulk sample measured at various points.

|  | Point A | Point B | Point C | Point D | Area scan |
|---|---|---|---|---|---|
| **Si/ at. %** | 50.47 (35) | 49.90 (35) | 50.55 (35) | 49.92 (35) | 50.31 (0) |
| **Ge/ at. %** | 39.57 (54) | 40.21 (54) | 38.98 (54) | 39.98 (54) | 40.06 (0) |
| **P/ at. %** | 8.25 (18) | 8.15 (18) | 8.55 (18) | 8.17 (18) | 7.85 (0) |
| **Fe/ at. %** | 0.61 (1) | 0.62 (1) | 0.61 (1) | 0.63 (1) | 0.63 (0) |
| **O/ at. %** | 1.09 (11) | 1.13 (11) | 1.31 (11) | 1.30 (11) | 1.16 (0) |

The SEM image and elemental mapping (**Figure S3**) for the preheated (1173 K) bulk sample shows inhomogeneity in P and Fe. The precipitates (highlighted in the dotted circles) were likely the same in size as that observed in XRD.

The EDX analysis and elemental mapping of the sintered sample show a very homogeneous distribution of elements over the whole sample to show a uniform gray color contrast for a very dense microstructure in SEM images. The composition was analyzed at various positions indicated by A, B, C, and D. The average composition measured for this sample was $Si_{50.3}Ge_{40.1}P_{7.9}Fe_{0.6}O_{1.1}$; hence, we succeeded in obtaining a thermodynamically stable supersaturated solid-solution of Si-Ge-P with small amount of Fe incorporation through high-energy ball milling.

Here, we would like to emphasize that the Si and Ge are prone to get easily oxidized in the ball milling process, but in general, the oxides are expelled out in the conventional high-temperature (1300 K) sintering process. However, in the low-temperature sintering case, it was unmanageable to remove the oxides and to get the oxidation free powder. Therefore the controlled atmosphere synthesis approach with hydrogen & argon gases was essential for minimizing the oxidation to a negligible value ~ 1 %. We consider that this tiny amount of oxygen would not make any significant contribution to the transport properties.

In-depth analysis by TEM (**Figure 3**) was carried out to observe the nano-crystallites and the formation of amorphous phase nature of in the samples. The as-milled powder had nano-crystallites (~ 9 nm) and amorphous-like disorder in the various portion of the TEM image, this could explain the halo pattern and peak broadening observed in the XRD pattern. Similarly, the sintered sample also showed very small portions of a disordered structure at the grain boundaries. We speculate that this nanostructure could effectively contribute to reduce the lattice thermal conductivity by strongly scattering the mid to high-frequency phonons. The crystallite sizes observed in TEM images were also consistent with XRD data calculated by the Scherrer's equation.

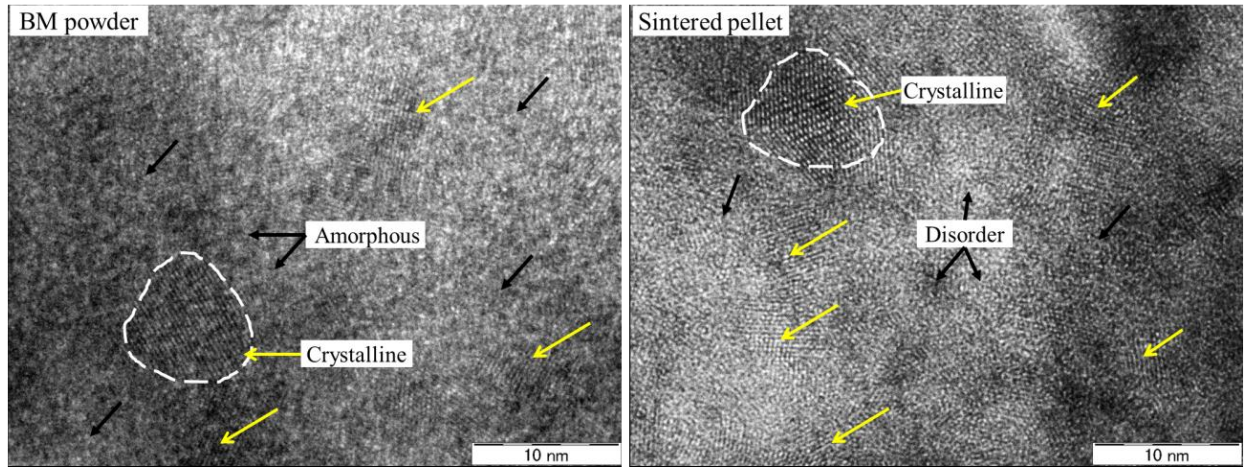

**Figure 3.** TEM image shows nanostructure of (a) ball-milled powder sample and (b) Sintered pellet. Yellow arrows are indicating crystalline region and black arrows showing disordered amorphous-like region.

TRANSPORT PROPERTIES

Sample preparation conditions such as milling time and sintering parameters are summarized in **Table 2**. Hereafter, reference sample, step (I + II), and step (I) samples are denoted as SM#1,[24] SM#2 and SM#3 throughout the manuscript.

**Table 2**. List of samples with their respective ball milling conditions, sintering parameters, and densities.

|  | Milling | Milling + Pause time / h | Temperature / K | Pressure / MPa | Soaking time / h | Density / gcm$^{-3}$ |
|---|---|---|---|---|---|---|
| SM 1 | Step (I + II) | 72 | 873 | 400 | 4.0 | 3.40 (5) |
| SM 2 | Step (I + II) | 72 | 873 | 400 | 4.0 | 3.32 (5) |
| SM 3 | Step (I) | 10 | 873 | 400 | 4.0 | 3.33 (5) |

A relatively large Seebeck coefficient $|S| > 300$ µV K$^{-1}$ (**Figure 4a**) was observed in the sample sintered after the alloying phase (SM#3). However, the absolute values were slightly smaller than the sample sintered after two-step milling (SM#1 and SM#2). All the transport properties were measured both during heating and cooling over the temperature range above and below the phase transition temperature.

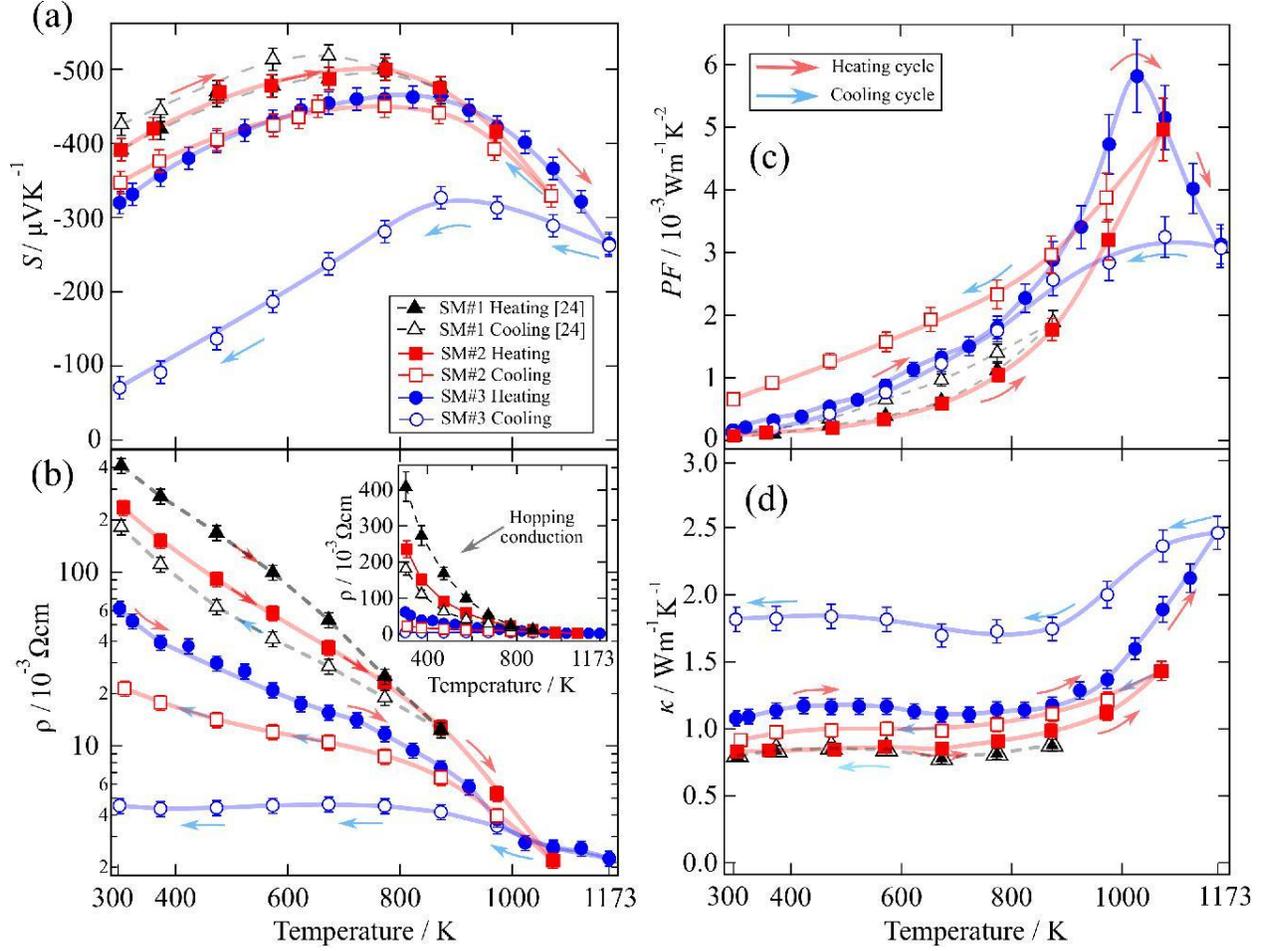

Figure 4. Temperature dependence of transport properties plotted for SM#1[24], SM#2, and SM#3 samples, (a) Seebeck coefficient, (b) electrical resistivity, (c) Power factor, and (d) thermal conductivity. The error bars were estimated from instrumental inaccuracies and variations from multiple measurements. The solid markers are for heating curves and empty markers for cooling curves. The red arrows indicate the direction for heating cycle and blue arrows for cooling cycle.

The Seebeck coefficients were extracted from the slopes of $\Delta V$ versus $\Delta T$ plots shown in **Figure S4**. The sign of the Seebeck coefficient was kept negative regardless of temperatures or samples. In the heating cycle for SM#3 heated up to 1073 K, the absolute value of Seebeck coefficient increased gradually to a maximum value of $|S| = 490 \pm 20$ μV K$^{-1}$ ($S = -490$ μV K$^{-1}$) up to 775 K, and the further increase of temperature made a drops of |S| down to a minimum of |S| = 330 ± 20 µV K$^{-1}$ (S = –330 µV K$^{-1}$) at 1173 K.

In the cooling cycle, the absolute value of Seebeck coefficient was dropped to lower values and reached a minimum of |S| = 330 ± 20 µVK$^{-1}$(S = –330 ± 20 µV K$^{-1}$) at 300 K from a maximum of |S| = 450 ± 20 µV K$^{-1}$ (S = –450 µV K$^{-1}$) at 873 K. Similar behavior was observed for SM#3 heated up to 1173 K, but with a significant degradation during the cooling cycle with least value of 70.70 ± 10 µV K$^{-1}$ at 300 K.

The thermal degradation in the supersaturated Si-Ge-Fe-P was uniquely observed for the first time in such compounds. The temperature dependence of electrical resistivity showed a finite difference in absolute values for the $\rho_{SM\#1}$, $\rho_{SM\#2}$, and $\rho_{SM\#3}$ samples are shown in the logarithmic plot of **Figure 4b**. The dissimilarity observed in the electrical resistivity must be due to the difference in the volume fraction of the nano-grains or amorphous-like microstructure caused by short and long ball milling condition. An exponential drop in electrical resistivity was observed as an increasing function of temperature shown in the inset of **Figure 4b**. While in the cooling cycle the resistivity showed very weak temperature dependence, which was almost a flat line in the linear scale. At room temperature, $\rho_{SM\#3}$ was 62.80 ± 0.6 mΩcm which monotonously decreased to 2.25 ± 0.2 mΩ cm at 1173 K, a drastic change by a factor of 28, but the $\rho_{SM\#3}$ increased while cooling by a factor of 2 to 4.5 mΩ cm at 300 K. The electrical resistivity behavior in heating-cooling cycle was reproduced in several samples to verify the reproducibility of this unusual phenomenon (**Figure S5**).

The estimated power factor ($S^2\sigma$) for SM#3 was slightly larger than the SM#1 throughout the temperature range (**Figure 4c**). The $PF_{SM\#3}$ increased gradually with increase in temperature and showed a huge peak of $PF_{SM\#3}$ = 5.8 ± 0.5 mWm$^{-1}$K$^{-2}$ at 1023 K, above this temperature, the $PF$ drops sharply to $PF_{SM\#3}$ = 3.1 ± 0.3 mW m$^{-1}$ K$^{-2}$ at 1173 K. In the cooling cycle, the peak for $PF_{SM\#3}$ was reduced by 50 %, and the heating & cooling curves were overlapped below 800 K.

The temperature dependence of thermal conductivities for $\kappa_{SM\#3}$, $\kappa_{SM\#2}$, and $\kappa_{SM\#1}$ were plotted in **Figure 4d**. A relatively smaller magnitude or a shift in the thermal conductivity was observed for the 60h milled-sintered ($\kappa_{SM\#1}$ and $\kappa_{SM\#2}$) compared to the 10h milled-sintered ($\kappa_{SM\#1}$) sample. The finite difference in $\kappa$ values for both the samples could be due to the volume fraction of nanostructure formation in the milling process. Effect of phase degradation was observed in the $\kappa_{SM\#2}$ and $\kappa_{SM\#3}$ behavior while heating above the transition temperature. In the heating, $\kappa_{SM\#3}$ starts at a very low magnitude of ~1 W m$^{-1}$ K$^{-1}$ with weak temperature dependence up to 800 K, while it increases monotonously from 1.2 ± 0.05 W m$^{-1}$ K$^{-1}$ (873 K) to 2.5 ± 0.07 Wm$^{-1}$K$^{-1}$ (1173 K). In the cooling cycle, $\kappa_{SM\#3}$ decreases gradually from 1173 K to 873 K (1.75 ± 0.05 W m$^{-1}$ K$^{-1}$) and stays uniform up to the room temperature (1.8 ± 0.05 W m$^{-1}$ K$^{-1}$) but throughout the temperature range $\kappa_{SM\#3}$ values in the cooling cycle stayed relatively larger than the heating cycle.

ACHIEVING FIGURE OF MERIT $ZT$ > 3.6

The temperature dependence of figure-of-merit $ZT$ was estimated for both heating & cooling cycles from the $S$, $\rho$, and $\kappa$. Results from this work was compared with the heavily doped nanostructured Si-Ge alloys reported by Bathula *et al.* and Wang *et al.* prepared by a similar route of high-energy ball milling (**Figure 5a**). In the heating cycle, the 10h milled-sintered (SM#3) showed a gradual increment in $ZT$, and a good consistency with the 60h milled-sintered (SM#1 & SM#2) reference data up to 873 K. Heating above the phase transition temperature (920 K), ZT of this sample exceeded 3 and a peak was observed at 1073 K with a huge value of $ZT_{SM\#3}$ = 3.70 ± 0.15. This value is definitely the highest ever reported or observed for bulk thermoelectric materials.

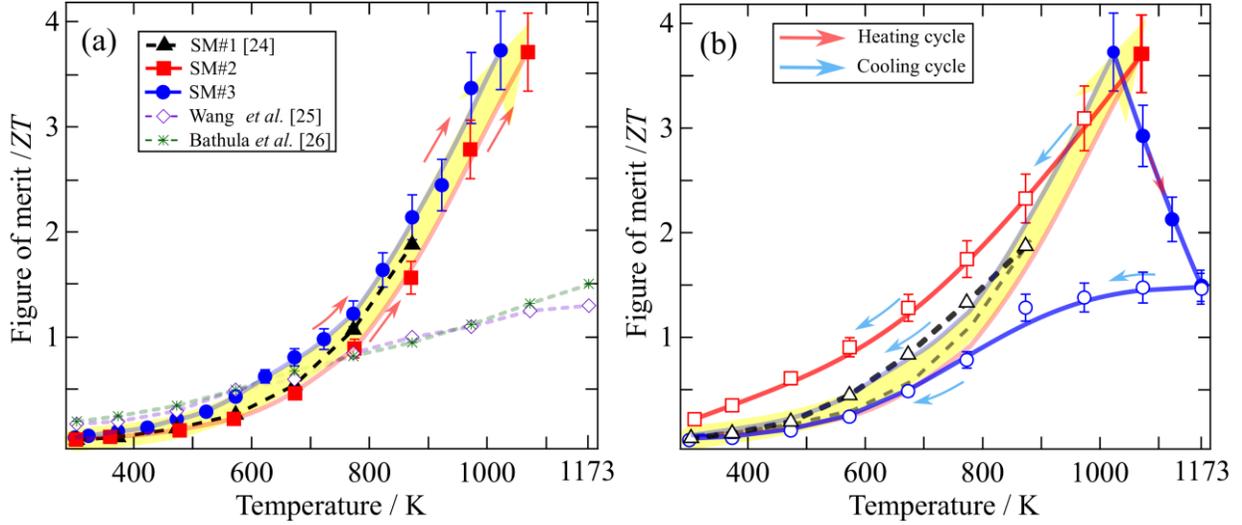

**Figure 5.** Temperature dependence of ZT (a) Measurements for supersaturated Si-Ge-Fe-P are compared with nanostructured reference samples in heating process,[25,26] (b) Heating – cooling cycles performed to observe the thermal stability. Multiple samples showing reproducibility of ZT > 3.6.

Further heating, the $ZT_{SM\#3}$ drops sharply to 1.50 ± 0.01 at 1173 K, as similar to the $PF_{SM\#3}$, did not recover back and gradually decreased with the cooling cycle (**Figure 5b**). Notably, the cooling data was in good agreement with the high ZT ~ 1.5 reported by Bathula *et al*. and Wang *et al*. The ZT was more than unity over a very wide temperature range from 800 K to 1173 K, and ZT > 2 in 900 K < T < 1123 K. In terms of applicability for commercial thermoelectric devices, possessing a high ZT > 1 in a wide temperature range is an extremely important and impressive result for practical applications.

DISCUSSION

The absolute values and the temperature-dependent behavior of the transport properties ($S$, $\rho$, $\kappa$) in the heating-cooling cycles were related to the three critical parameters, i.e. (a) volume fraction of the amorphous-like or disordered nanostructure, (b) Fe incorporation in supersaturated phase, and (c) effect of precipitation or phase transformation. The amorphous-like nanostructure produced by the high-energy ball milling possesses microstructural defects such as point defects, line defects/ dislocations, grain boundaries responsible for scattering phonons and carriers. Thus, affecting the absolute values of the electrical resistivity and thermal conductivity.

The higher Seebeck coefficient was in good agreement with the presumably formed impurity peak of Fe 3$d$ at the conduction band edge, while the difference in the synthesis condition, i.e., 10 h milling and 60 h milling could cause a difference in the chemical composition. In the EDX composition showed a finite difference in the Fe incorporation 0.6 at.% and 0.9 at.% for short (SM#3) and long (SM#1) milling sample, respectively. The above results hint a study on Fe concentration-dependent transport properties that could be done in the near future and would be published elsewhere.

The $\rho_{SM\#3} < \rho_{SM\#1}$ over the temperature range should be due to a difference in the volume fraction of amorphous-like disorder. Slightly larger crystallite or grain sizes (15.80 ± 0.5 nm) for SM#3 compared to 13 ± 0.5 nm (SM#1) could represent better crystallinity, which means the formation of a superior conduction path for the charge carriers. A distinctive negative temperature coefficient of resistivity (TCR) was observed in the heating cycle for all the samples. The exponential decay in electrical resistivity could be misinterpreted for semi-conductive nature caused by electron – hole excitation. We strongly deny the contribution from electron – hole excitation in electrical resistivity because the supersaturated phases were heavily doped with large amount of 10 at.% P. Along with, Seebeck coefficient shows increasing tendency with the temperature, and thermal conductivity shows weak temperature dependence. The bipolar-diffusion range (the temperature range of electron – hole excitation) was 800 – 1173

K, where the degradation in transport properties, that are a decrease of |S| and increase of |$\kappa_{el}$|, was observed.

Another possible mechanism for the negative TCR is hopping conduction, which are known to possess similar negative TCR. The origin of hopping conduction is related to the hopping of carriers with thermal excitation energy from one to another localized-state. The localized states are formed in the bandgap of the disordered or amorphous semiconductors.[60–68] The hopping of the carriers to the nearest neighbor empty site is the Nearest-Neighbor Hopping conduction (NNH) (ln$\sigma \propto T^{-1}$) or hopping of carriers to the next energy level irrespective of spatial distribution is Variable Range Hopping (VRH) conduction. There are two types of VRH conduction (a) 3D Mott VRH (ln$\sigma \propto T^{-1/4}$)[69] and (b) Efros-Shklovskii / ES-VRH (ln$\sigma \propto T^{-1/4}$).[70] For NNH, we observed ln$\sigma$ vs. 1000/T two components (**Figure 6a**), one at low-temperature with activation energy 60 – 70 ± 10 meV and other at high-temperature 300 – 600 ± 20 meV. Similar linear fitting with two components was observed with 3D-Mott-VRH (**Figure 6b**) and ES-VRH (**Figure 6c**). This analysis strongly indicates that the hopping conduction with one or a combination of the mechanisms described above occurred in the supersaturated solid-solution of Si-Ge-Fe-P.

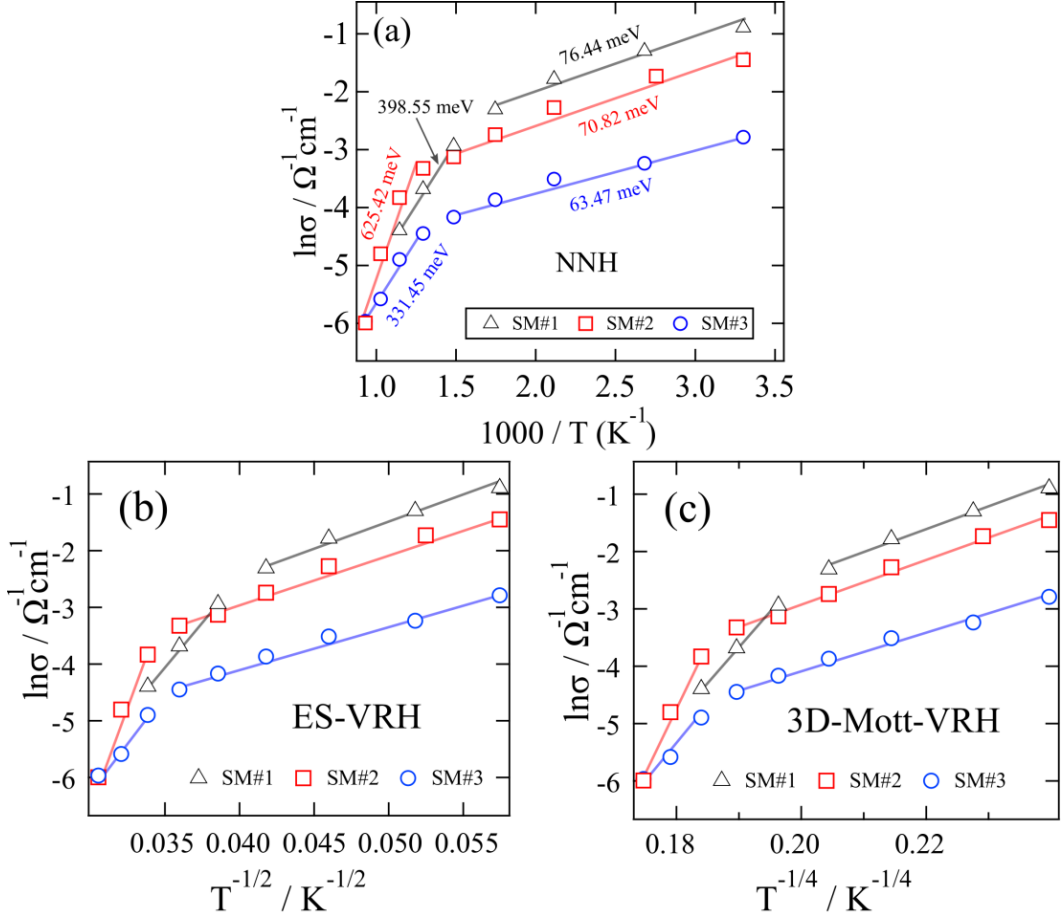

**Figure 6.** (a) $\ln\sigma$ versus $T^{-1}$ plot for nearest-neighbor hopping (NNH) regime with estimated activation energies ($E_{NNH}$), (b) $\ln\sigma$ versus $T^{-0.5}$ plot for 3D-Efros-Shklovskii variable-range hopping (ES-VRH) conduction, and (c) $\ln\sigma$ versus $T^{-0.25}$ plots for 3D-Mott variable-range hopping (3D-Mott-VRH) conduction.

The hopping conduction was not observed in cooling data mainly due to loss of localized states formed in strong relation with the disappeared amorphous-like microstructure. The Seebeck coefficient, on the other hand, was drastically degraded due to the change in the number of Fe impurity states and the carrier concentration in association with the precipitation of secondary phase. This fact was in good agreement with the previously reported study[24] in which the precipitated sample showed less amount of Fe incorporated in the main phase that destructs the effective modification of the electronic density of states near the chemical potential. The precipitation could be similar to the studies reported for dopant (phosphorus) precipitation in n-type Si-Ge alloys.[71–76] As a result, the power factor was reduced by simultaneous alterations in the electrical resistivity and Seebeck coefficient post – phase transition temperature. Damodara Das *et al.* and Li *et al.* reported similar irreversible amorphous – crystalline transition for amorphous $Se_{20}Te_{80}$/$Sb_2Te_3$ thin films[77–79] and Uenishi *et al.* for Ag-Cu super-saturated solid solution.[80] Such a transition in heating-cooling cycles resulted in degradation of Seebeck coefficient and electrical resistivity. All the evidence supports our argument of formation of amorphous-like supersaturated solid-solution which transforms through crystallization and precipitation above certain transition temperature.

The thermal conductivities for all the samples were remarkably low ($\kappa \sim 1$ $Wm^{-1}K^{-1}$), which must be due to the nanostructuring. The obtained amorphous-like nanostructure has immeasurable interfaces (grain boundaries, point defects, dislocations, and alloying effect), which could effectively scatter a wide range of phonons (low-high frequency)[81]. Such substantial phonon scattering resulted in shorter phonon mean free paths. The 20 % reduction in thermal conductivity for $\kappa_{SM\#1}$ and $\kappa_{SM\#2}$ compared to $\kappa_{SM\#3}$ should be due to additional defects induced in the grinding phase.

We also suspect some additional mechanisms such as electron-phonon interactions, which might have involved in obtaining such low-thermal conductivity. The model for reducing thermal conductivity in heavily doped ($10^{19} - 10^{21}$ $cm^{-3}$) degenerate semiconductors was postulated by Vining *et al.*,[82] Liao *et al.*,[83]. Assuming the above models, experimentally, reduction in lattice thermal conductivity was proposed for heavily doped Si and Si-Ge alloys.[84,85] The supersaturated Si-Ge-Fe-P phases satisfy above electron – phonon interaction criteria as it contains a large amount of P (10 at. %) with a carrier concentration of $\sim 10^{19}$ $cm^{-3}$.[24]

In the temperature range from 900 K to 1173 K ($\kappa_{SM\#3}$), the bipolar diffusion, thermal excitation, and the degradation of supersaturated solid-solution (increased crystallite size and formation of precipitates) assisting phonon transport. These conditions resulted in higher thermal conductivity. Although the cooling curve showed 80 % deviation in the absolute value of $\kappa$ was < 2 W m$^{-1}$ K$^{-1}$, which was still low, indicating a presence or retention of nanostructures and probable phonon scattering at the precipitates.

Indeed, the obtained figure-of-merit $ZT$ = 3.70 ± 0.15 is an overwhelming value achieved in the metastable supersaturated solid-solution of Si-Ge-Fe-P. The future work finding out a plausible solution to the thermal degradation problem should be of great interest. It is worth mentioning that there are several key factors; ball to powder ratio, ball size, milling time, milling speed, sintering temperature, sintering pressure, soaking time, Ar/H ratio, oxidation, and heating − cooling rate in measurement that are responsible for reproducibility of the sample composition as well as the transport properties.

CONCLUSION

We succeeded in obtaining a very low thermal conductivity (1 ± 0.15 W m$^{-1}$ K$^{-1}$) in a nanostructured (10 ± 0.5 nm) supersaturated Si-Ge-P-Fe compounds by inducing strong phonon scattering mechanisms. We confirmed that the Fe impurity state improves the Seebeck coefficient (> 450 µV K$^{-1}$) by measuring the transport property through a phase transition. An overwhelming figure-of-merit $ZT$ = 3.70 ± 0.15 at 1073 was obtained in the supersaturated form, which unfortunately degraded under the effect of secondary phase formation. Reproducibility and thermal stability of such impressive $ZT$ were also confirmed in the supersaturated phases. A thorough investigation on Fe and P concentration dependence of transport properties, cost-effective high-performance with low Ge concentration, and Ge free supersaturated solid-solutions of silicon nanostructures with advanced spectroscopy techniques will be of great interest and investigated in the near future.

ASSOCIATED CONTENT

Supporting Information.

XRD analysis of milled powder and sintered pellets. SEM and EDX images of ball-milled powder and sintered sample. Raw data of Seebeck measurements and temperature dependence of electrical resistivity in cyclic heating-cooling cycles.

Acknowledgment

This work was conducted under the financial support of JST CREST and Japan Society for the Promotion of Science (JSPS) KAKENHI, Grant Nos. 18H01695, 18K18961.

**Table of Contents (ToC) / Abstract Graphic**

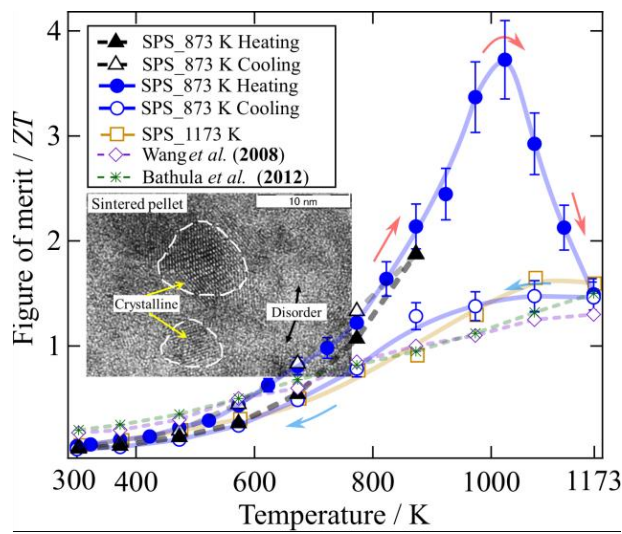

# Supporting Information

# Distinctive Thermoelectric Properties of Supersaturated Si-Ge-P Compounds: Achieving Figure of Merit ZT > 3.6

*Swapnil Ghodke,*[†,*] *Omprakash Muthusamy,*[†,*] *Kévin Delime–Codrin,*[†] *Seongho Choi,*[†] *Saurabh Singh,*[†] *Dogyun Byeon,*[†] *Masahiro Adachi,*[‡] *Makoto Kiyama,*[‡] *Takashi Matsuura,*[‡] *Yoshiyuki Yamamoto,*[‡] *Masaharu Matsunami,*[†] *Tsunehiro Takeuchi*[†]

[†] Energy Materials Laboratory, Toyota Technological Institute, 2-12-1 Hisakata, Tempaku Ward, Nagoya, 468-0034, Japan

[‡]Sumitomo Electric Industries Ltd., Itami, Hyogo 664-0016, Japan.

*E-mail: swapneelghodke@gmail.com, omprakashmuthusamy@gmail.com

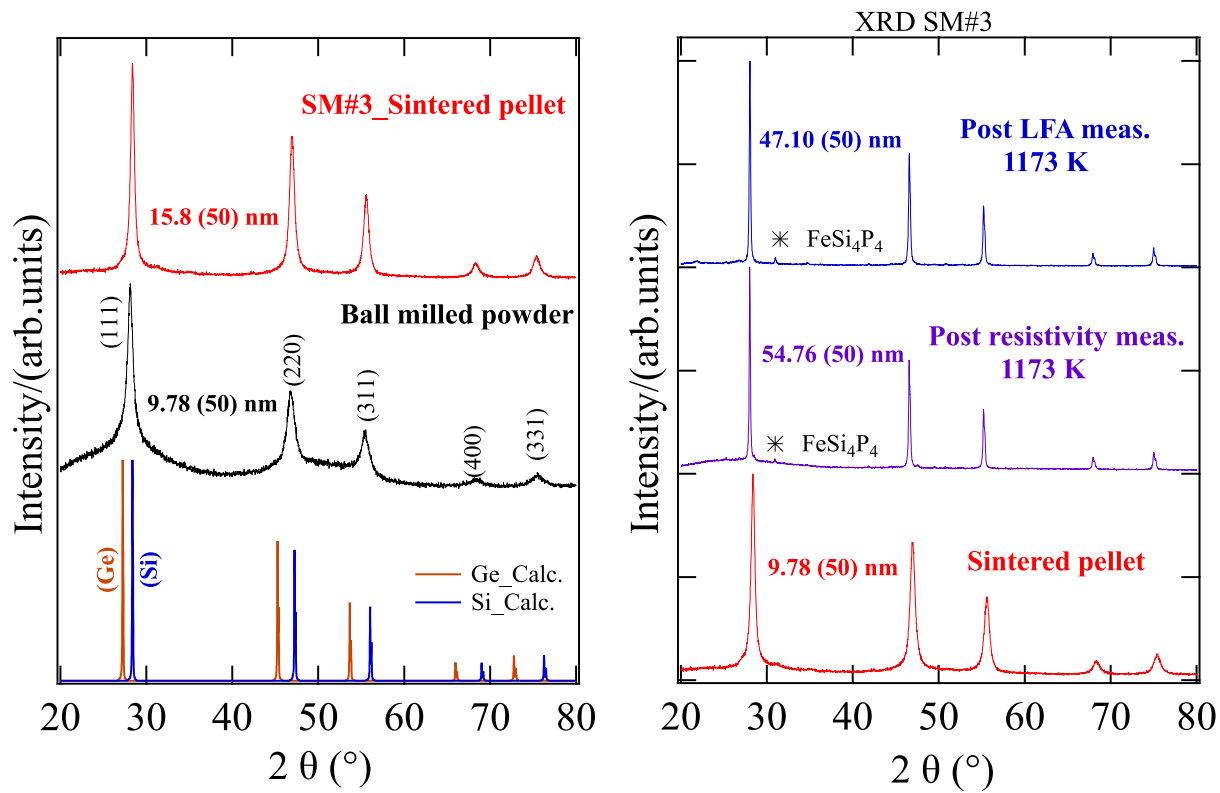

**Figure S1.** (Left) XRD patterns of ball milled powder, sintered pellet, and calculated Si and Ge. (Right) XRD patterns measured at for as-sintered pellet, post resistivity measurement and post LFA measurement. In resistivity and LFA measurements samples were heated up to 1173 K.

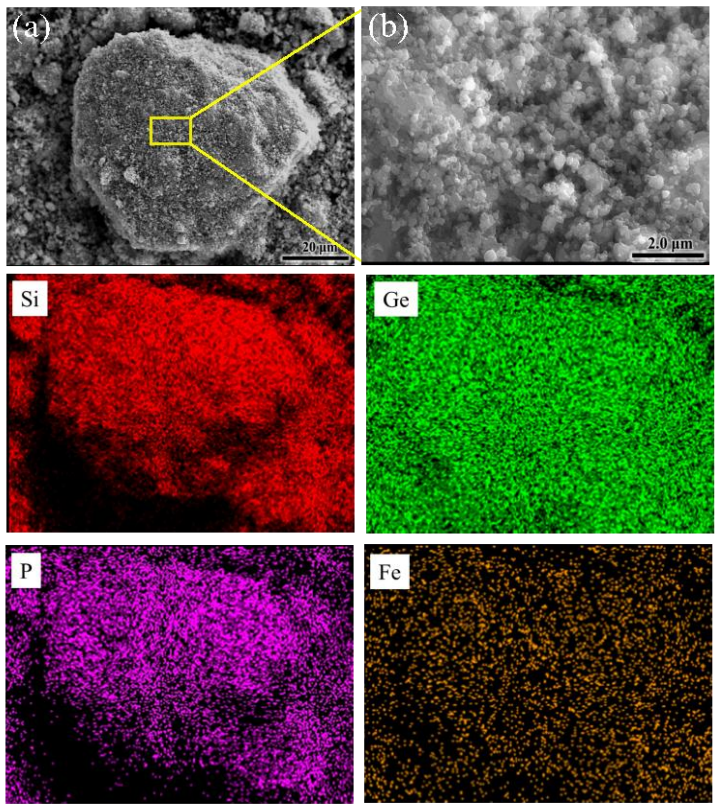

**Figure S2.** Typical SEM images of ball-milled powder (a) low-magnification image and (b) magnified portion of the agglomerated fine particles. Elemental mapping (bottom) for Si, Ge, P, and Fe elements.

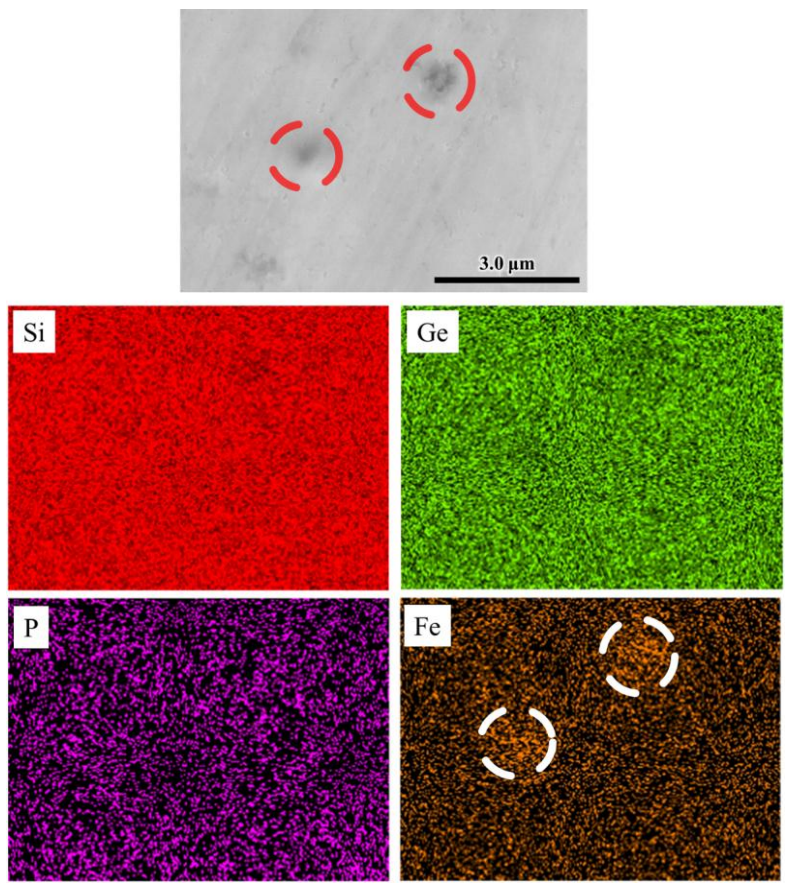

**Figure S3.** SEM image of the bulk sample (SM#3) measured post resistivity measurement. The circles indicate precipitation of impurity phases. Elemental mapping (bottom) for Si, Ge, P, and Fe elements.

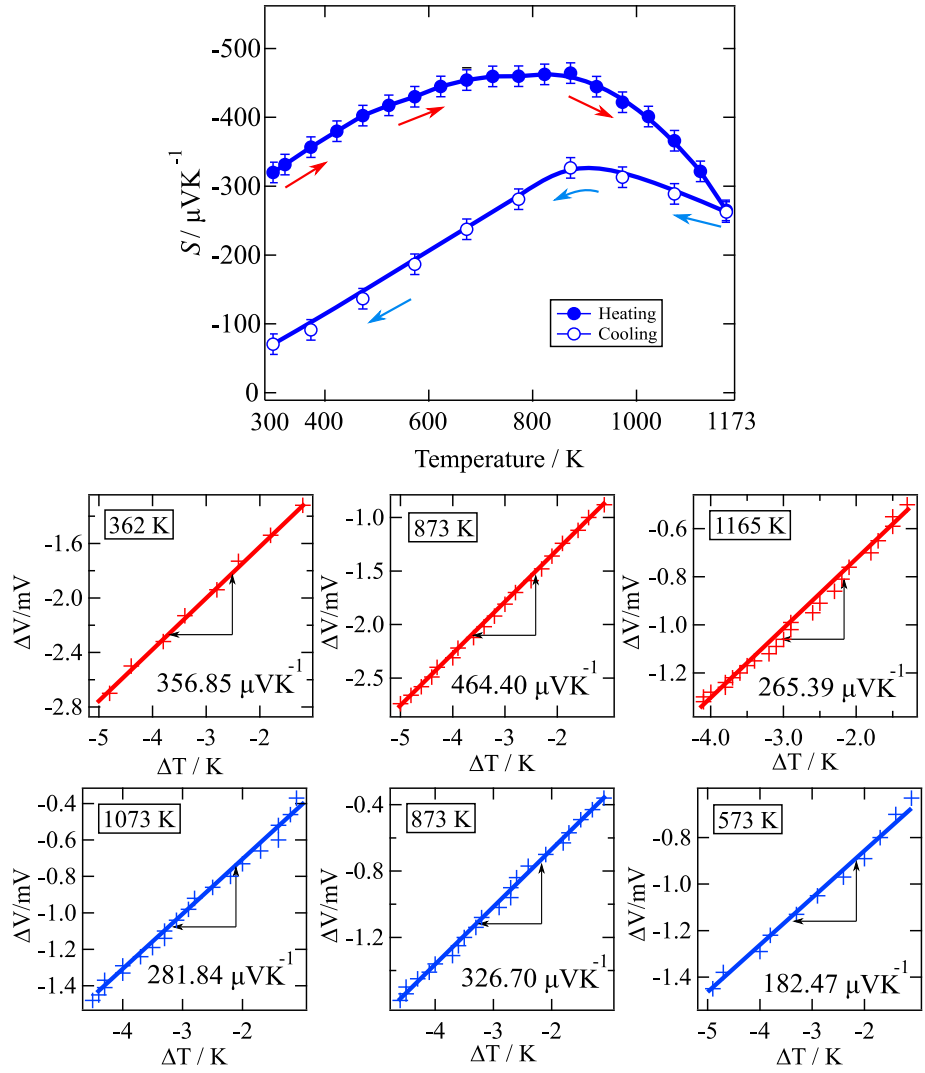

**Figure S4.** (top) Typical temperature dependence of Seebeck coefficient ($S_{SM\#3}$) measured by steady-state method. (bottom) Raw data of the ΔV versus ΔT plots obtained at selected temperatures while heating and cooling cycle.

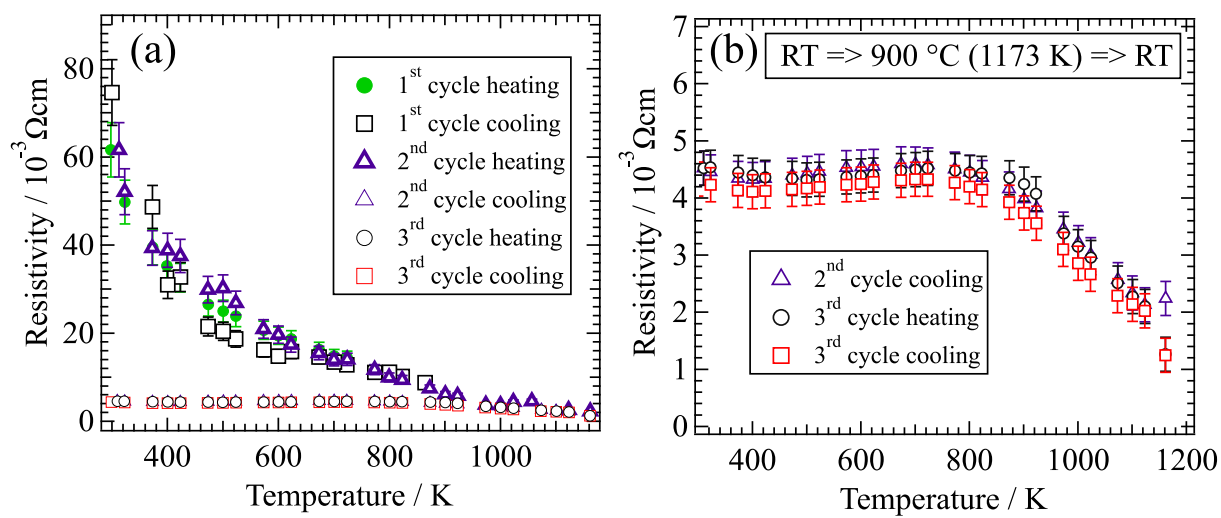

**Figure S5.** Temperature dependence of electrical resistivity plotted for SM#3 sample, (a) 1st cycle was measured from 300 – 873 K, 2nd cycle from 300 – 1173 K, and 3rd cycle 300 – 1173 K. (b) Magnified data plot for 2nd and 3rd cycle. All the thermal cycles were performed in succession on the same (piece) sample. The error bars were estimated from instrumental inaccuracies and variations from multiple measurements.